\documentclass[prb,floatfix,twocolumn,notitlepage,showpacs,superscriptaddress]{revtex4-1}
\usepackage{amssymb}
\usepackage{graphicx}
\usepackage{epstopdf}
\usepackage{float}
\usepackage[caption=false]{subfig}

\usepackage{amsfonts,amsmath} \usepackage{bm} \usepackage{dcolumn}
\usepackage{epsfig} \usepackage{latexsym}
\usepackage{graphicx}
\usepackage{multirow}

\usepackage{graphicx}
\usepackage{epsfig}
\usepackage[usenames]{xcolor}
\usepackage{amsfonts,amsmath}
\usepackage{bm}
\usepackage{dcolumn}
\usepackage{hyperref}

\newcommand{\<}{\langle}
\renewcommand{\>}{\rangle}

\usepackage{color}

\newcolumntype{M}[1]{>{\centering\arraybackslash}m{#1}}

\begin{document}
\title{Nature of a single doped hole in two-leg Hubbard and $t$-$J$ ladders}
\author{Shenxiu Liu}
\affiliation{Department of Physics, Stanford University, Stanford, CA 94305, USA}
\affiliation{Stanford Institute for Materials and Energy Sciences,
SLAC National Accelerator Laboratory and Stanford University, Menlo Park, CA 94025, USA}
\author{Hong-Chen Jiang}
\affiliation{Stanford Institute for Materials and Energy Sciences, SLAC National Accelerator Laboratory and Stanford University, Menlo Park, CA 94025, USA}
\author{Thomas P. Devereaux}
\affiliation{Stanford Institute for Materials and Energy Sciences,
SLAC National Accelerator Laboratory and Stanford University, Menlo Park, CA 94025, USA}
\affiliation{Geballe Laboratory for Advanced Materials, Departments of Physics and Applied Physics, Stanford University, Stanford, California 94305, USA}
\date{\today}

\begin{abstract}
In this paper, we have systematically studied the single hole problem in two-leg Hubbard and $t$-$J$ ladders by large-scale density-matrix renormalization group calculations. We found that the doped hole in both models behaves similarly with each other while the three-site correlated hopping term is not important in determining the ground state properties. For more insights, we have also calculated the elementary excitations, i.e., the energy gaps to the excited states of the system. In the strong rung limit, we found that the doped hole behaves as a Bloch quasiparticle in both systems where the spin and charge of the doped hole are tightly bound together. In the isotropic limit, while the hole still behaves like a quasiparticle in the long-wavelength limit, its spin and charge components are only loosely bound together with a nontrivial mutual statistics inside the quasiparticle. Our results show that this mutual statistics can lead to an important residual effect which dramatically changes the local structure of the ground state wavefunction.
\end{abstract}
\maketitle

It is believed that a minimal model which captures the strong-correlation physics of high-temperature superconductivity is the Hubbard model and its strong-coupling limit, the $t$-$J$ model. Of fundamental importance is the question of whether these simple models contain the basic requirements to support superconductivity in the phase diagram as compared to other phases.\cite{Lee2006RMP,Zhang1988,Weng1996,White1997,White1998,Fradkin2012NaturePhys,Fradkin2015RMP,Liu2012PRL} However, despite decades of studies, the precise nature of these simple models are still not well understood in two dimensions (2D), including the ground state properties. Alternatively, ladder systems serve as a bridge from one-dimensional (1D) chains to 2D systems, offering a rich playground for studying the interplay of charge and spin degrees of freedom and providing a pathway to understand the physics of strongly correlated systems. Specifically, the motion of single hole doped into the two-leg ladder antiferromagnet is a fundamental issue to start with, where the key physics of the problem is the competition between the antiferromagnetic correlation and the kinetic energy of the hole.

Theoretically, it has long been thought that the undoped two-leg $t$-$J$ ladder is adiabatically related to a band insulator, and earlier numerical studies supported the idea that the doped holes form conventional Bloch quasiparticles.\cite{Schmitt1988,Kane1989,Martinez1991,Dagotto1994,Leung1995} Quite strikingly, recent large-scale density-matrix renormalization group (DMRG) studies provide strong evidences to show that a single hole doped to the two-leg $t$-$J$ ladder localizes at large length scales, breaking the translational symmetry, which is sharply incompatible with Bloch's theorem for any quasiparticle state\cite{Weng1996,Weng2001,Zhu2013TJ,Zhu2015Charge,Zhu2015Quasiparticle}. A more recent DMRG study of the same model however reached an opposite conclusion where the hole behaves as a conventional quasiparticle with finite quasiparticle spectral weight, and there is no charge localization\cite{White2015TJ}. As a result, this seemingly simple problem is still in debate, requiring further independent investigation.

Besides the $t$-$J$ model, it is also important to study the Hubbard model directly to investigate the properties of a single doped hole. This is due to three-site correlated hopping terms discarded in traditional $t$-$J$ model studies for simplicity. Indeed, this three-site correlated hopping term was shown to be crucial to understand the origin of the strongly dispersive feature found at high binding energy in the spectral function of the Hubbard model\cite{WangYao2015}, as well as contributing to the persistence of spin excitations with hole doping and hardening of spin excitations with electron doping in the Hubbard model\cite{Jia2014NatureComm}. A natural question is that whether these three-site correlated hopping terms still play an important role in the ground state properties of the single hole.

\begin{figure}[htbp]
  \includegraphics[width=\linewidth]{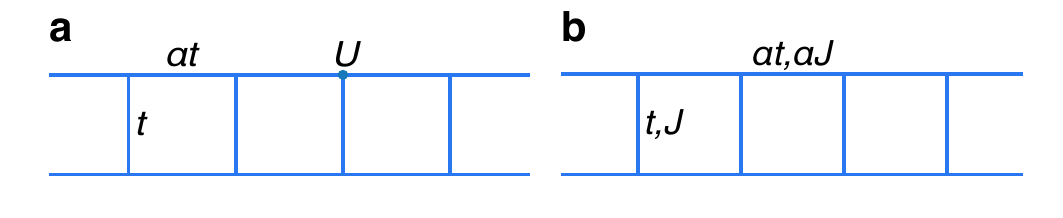}
\caption{(Color online) The parameters of the (a) anisotropic Hubbard model and (b) $t$-$J$ model on a two-leg square ladder. Here $t_{ij}=t$ ($t_{ij}=\alpha t$) and $J_{ij}=J$ ($J_{ij}=\alpha J$) describe the inter-chain (intra-chain) hopping and spin superexchange couplings, respectively. At $\alpha=1$, it reduces to the isotropic limit.}\label{Fig:Model}
\end{figure}

\begin{figure}[!htbp]
  \centerline{\includegraphics[width=\linewidth]{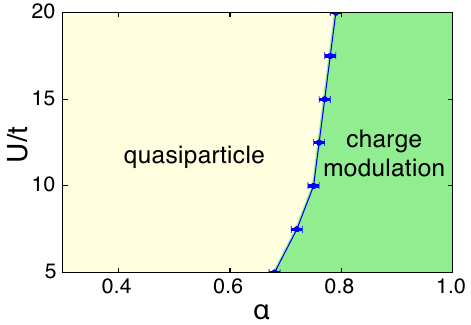}}
  \caption{(Color online) Ground state phase diagram of the two-leg Hubbard ladder. The blue line is the phase boundary labeled by the critical value $\alpha_c$ as a function of $U/t$. In the left region ($\alpha<\alpha_c$), there is no charge modulation in the hole density profile, while the right region ($\alpha>\alpha_c$) has clear charge modulation.}\label{Fig:HubbardPhase}
\end{figure}

To answer these questions, we will focus on the two-leg Hubbard and $t$-$J$ ladders in this paper in which the undoped spin background remains gapped. We have carried out systematic DMRG \cite{White1992DMRG} simulations to extract the ground state properties of ladders of length up to $L=300$ and elementary excitations of a single doped hole to obtain unprecedentedly complete information. Following previous studies\cite{Zhu2013TJ,Zhu2015Charge,Zhu2015Quasiparticle,White2015TJ}, we have considered a range of values of the parameter $\alpha$, the ratio of the hopping matrix elements and the spin exchange couplings on the legs and the rungs of the ladders of the Hubbard and $t$-$J$ models, see Fig. \ref{Fig:Model}. Our principle results are:%

(1) In contrast to the high energy spectrum, we find that the single hole doped in the two-leg Hubbard ladder behaves similarly with the $t$-$J$ model as shown in the phase diagram in Fig. \ref{Fig:HubbardPhase}. In the strong rung limit ($\alpha<\alpha_c$), the injected hole behaves as a Bloch quasiparticle and there is no charge modulation in the hole density profile. As the ladder anisotropic parameter is continuously tuned from the strong rung limit beyond a critical value $\alpha_c$ towards the isotropic limit ($\alpha=1$), a strong charge modulation appears in the hole density distribution. Therefore, we conclude that the three-site correlated hopping term is not crucial in determining the ground state properties of the single doped hole.%

(2) We found that the elementary excitation energy of the single doped hole in two-leg $t$-$J$ ladder scales as $1/L^2$ in both limits, including the strong-rung limit $\alpha<\alpha_c$ and isotropic case $\alpha=1$, supporting the quasiparticle behavior of the doped hole. As the single hole doped in both models has similar behavior, our conclusion for the $t$-$J$ ladder should also apply for the Hubbard ladder, i.e., the injected hole also behave as a quasiparticle in the isotropic limit.%

(3) Although in the long-wavelength limit, the single doped hole behaves similarly in both the strong rung limit ($\alpha<\alpha_c$) and isotropic case ($\alpha=1$), there is a significant difference between them. The spin and charge of the hole are tightly bound together as a quasiparticle at a length scale $\xi\sim 1$ lattice spacing in the strong-rung limit, however, they are only loosely bound together at a longer length scale $\xi\sim 3$ lattice spacings  in the isotropic case. This loose quasiparticle has a nontrivial internal structure since the spin partner can now move away from the charge partner, leading to an important residual effect which significantly changes the local structure of the ground state wavefunction.

In section \ref{Sec:Model}, we define the model Hamiltonians of the two-leg $t$-$J$ and Hubbard ladders. In section \ref{Sec:Hubbard}, we present our DMRG results for the two-leg Hubbard ladder, and calculate the elementary excitations of the single hole doped in the two-leg $t$-$J$ ladder in section \ref{Sec:Elementary}. Section \ref{Sec:ResidualEffect} is devoted to the residual effect and a conclusion is given in section \ref{Sec:Conclusion}.

\section{Model Hamiltonian} \label{Sec:Model}

The anisotropic Hubbard model on the two-leg square ladder is defined as\cite{Zhu2016Hubbard}
\begin{eqnarray}
H_{H}=-\sum_{\langle ij\rangle \sigma} t_{ij} \left(c_{i\sigma}^\dagger c_{j\sigma} + \mathrm{h.c.}\right) + U\sum_i n_{i\uparrow}n_{i\downarrow}, \label{Eq:Hubbard}
\end{eqnarray}
where $\<ij\>$ indicates nearest-neighbor (NN) bonds with hopping integral $t_{ij}=t$ on the rungs and $t_{ij}=\alpha t$ on the legs, as sketched in Fig. \ref{Fig:Model}(a). $c_{i\sigma}^\dagger$ creates an electron on site $i$ with spin polarization $\sigma$. The electron number operator is $n_i=\sum_\sigma c_{i\sigma}^\dagger c_{i\sigma}$, and $U$ is the on-site repulsion.

The $t$-$J$ Hamiltonian on two-leg ladder is given by (see Fig. \ref{Fig:Model}(b))
\begin{equation}
H_{tJ}=-\sum_{\< ij\>\sigma} t_{ij}\left(c_{i\sigma}^\dagger c_{j\sigma} + \mathrm{h.c.}\right) +  \sum_{\< ij\>} J_{ij}\left(\mathbf{S}_i\cdot\mathbf{S}_j - \frac{1}{4}n_in_j\right). \label{Eq:TJMODEL}
\end{equation}
Similar with the Hubbard ladder, $t_{ij}=t$ labels the hopping integral on the rungs and $t_{ij}=\alpha t$ on the legs. The spin superexchange interactions on the rungs and legs are given by $J_{ij}=J$ and $J_{ij}=\alpha J$, respectively. $\mathbf{S}_i$ is the spin operator on site $i$. Different than the Hubbard model, the action of the $t$-$J$ Hamiltonian is restricted to the Hilbert space constrained by the no-double-occupancy condition, i.e., the number operator $n_i\leq 1$. The site  index $i=(x,y)$ with $y=1,2$ denoting the two legs and $x$ runs from $1$ to $L$. Following the previous studies\cite{Zhu2013TJ,Zhu2015Charge,White2015TJ}, we consider the same range of parameters $0<\alpha\leq 1$ for both models .\footnote{Note that this is not the exact $t$-$J$ correspondence of the Hubbard model we are studying, which should be $J_{ij} = \alpha^2J$ on rungs. However, this doesn't lead to any qualitative difference.} Specifically, in the following DMRG calculation, we will set $t=1$ as an energy unit for the Hubbard ladder, while set $J=1$ as an energy unit for the $t$-$J$ ladder. 

\section{Single hole in the two-leg Hubbard ladder} \label{Sec:Hubbard}
Previous studies\cite{Zhu2013TJ,Zhu2015Charge,White2015TJ} of the two-leg anisotropic $t$-$J$ ladder show that a nontrivial charge modulation appears in the hole density profile in the isotropic limit $\alpha=1$, which is sharply different than the strong-rung limit $\alpha<\alpha_c$. However, whether this still holds true for the Hubbard model is unknown due to the presence of the three-site correlated hopping terms. To answer this question, we have performed an extensive DMRG study on the two-leg Hubbard ladder (see Eq. (\ref{Eq:Hubbard}) and Fig. \ref{Fig:Model}(a)). Our study shows that a single hole doped in the Hubbard ladder behaves similarly with that in the $t$-$J$ ladder, where the results are summaried in the phase diagram in Fig. \ref{Fig:HubbardPhase} for $5\leq U/t\leq 20$ and $0<\alpha \leq 1$. Specifically, there are two distinct phases, a conventional quasiparticle phase without charge modulation for $\alpha<\alpha_c$ and an interesting charge modulation phase for $\alpha>\alpha_c$. Actually, this charge modulation was shown to appear even on very small clusters suggesting its robustness.\cite{Zhu2016Hubbard}

\begin{figure}[!htbp]
  \centerline{\includegraphics[width=0.5\textwidth]{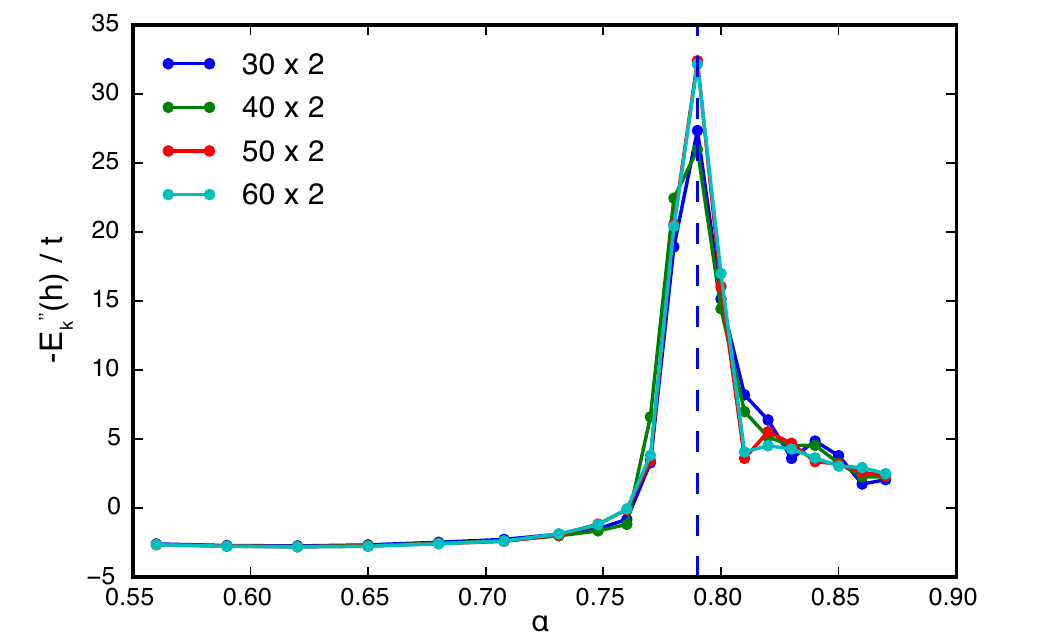}}
  \caption{(Color online) Second derivative of single hole kinetic energy $E_k''(h) = d^2E_k^h / d\alpha^2$ vs. $\alpha$ with different sample sizes at $U/t=20$. The peak position, labeled by the blue dashed line, determines the phase boundary between the two distinct phases.}\label{Fig:KineticEnergy}
\end{figure}

The phase diagram in Fig. \ref{Fig:HubbardPhase} is determined from standard DMRG simulations, where a sufficient large number of DMRG states (see below) were kept to limit the truncation error per step to $\leq 10^{-7}$. For each system size, the ground states at half-filling and with a single doped hole were accurately obtained. The phase boundary between the two distinct phases was determined by calculating the single hole kinetic energy $E_k^h$, compared to half-filling, as
\begin{eqnarray}
E_k^h = E_k^{\mathrm{one-hole}} - E_k^{\mathrm{half-fill}}.\label{Eq:HoleKinetic}
\end{eqnarray}
Here $E_k^{\mathrm{one-hole}}$ is the ground state kinetic energy of the system with one hole, and $E_k^{\mathrm{half-fill}}$ is the kinetic energy at half-filling. Therefore, $E_k^h$ solely represents the kinetic energy of the single injected hole. For a fixed $U/t$, the second derivative of the single hole kinetic energy, i.e., $E_k^{\prime\prime}(h)=\frac{d^2E_k^h}{d\alpha^2}$, shows a sharp peak at the critical value $\alpha=\alpha_c$, labeling the phase boundary between the two distinct phases. As an example, Fig. \ref{Fig:KineticEnergy} shows $E_k^{\prime\prime}(h)$ as a function of $\alpha$ for $U/t=20$ and various system sizes. It is clear that the finite-size effect is negligible and our results of $\alpha_c$ represent the reliable value in the thermodynamic limit, i.e., $L\to \infty$. In the following, we will directly compare the Hubbard ladder with the $t$-$J$ ladder in various aspects and provide evidences to show that a single hole doped in both models behaves similarly.

\begin{figure}[!htbp]
\includegraphics[width=\linewidth]{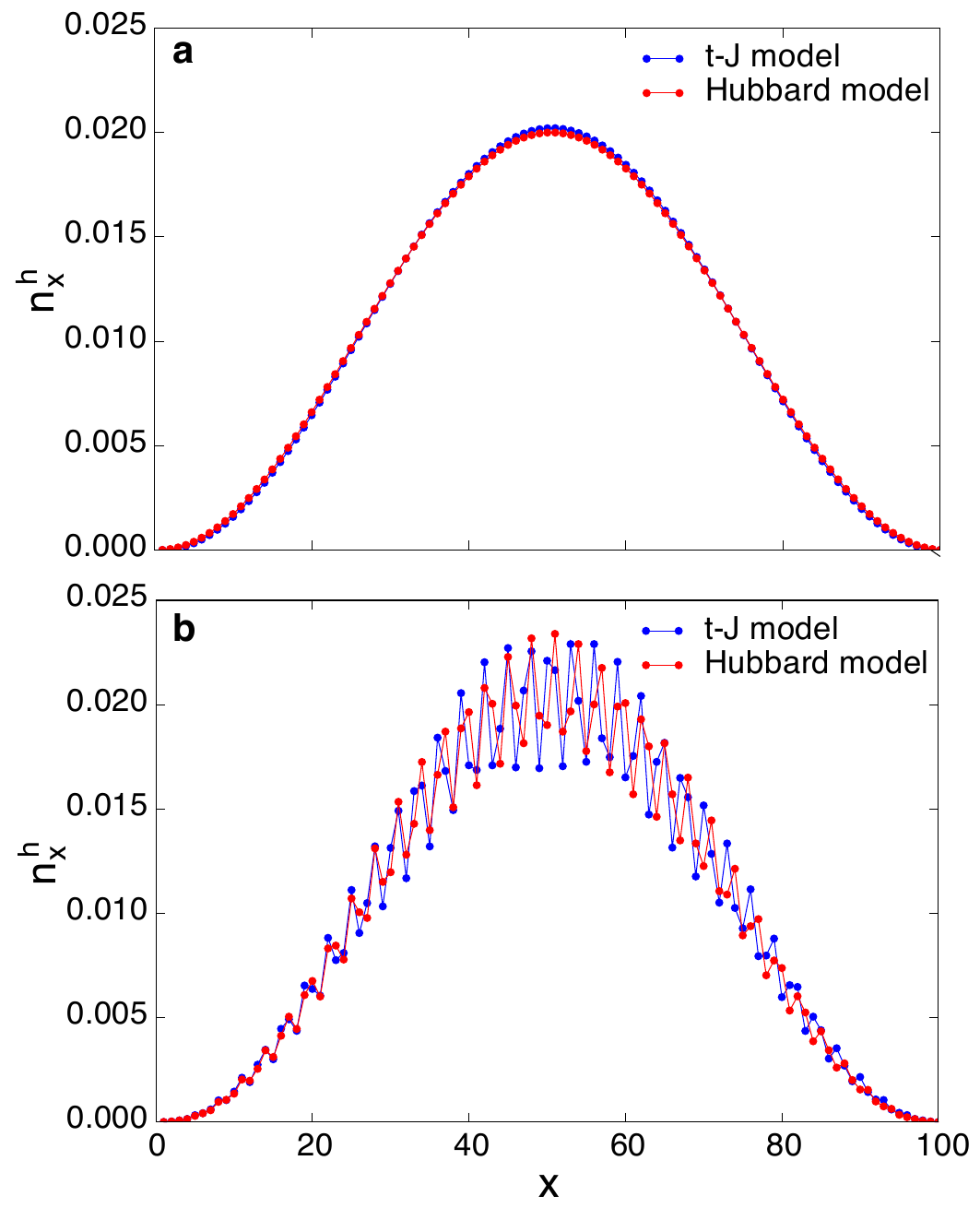}
\caption{(Color online) Hole density distribution $\left<n_x^h\right>$ of Hubbard model at $U/t=20$ and $t$-$J$ model at $t/J=5$ for (a) $\alpha=0.5$ and (b) $\alpha=1.0$. Here the system size is $100\times 2$ and $x$ is the rung index.}\label{Fig:tJHubSim}
\end{figure}

\paragraph{Hole density distribution:} %
We first calculate the hole density distribution function $\left<n_x^h\right> = \sum_{y} \left(1 - \left<n_{x,y}\right>\right)$ for both the Hubbard and $t$-$J$ ladders, where $x$ is the rung index and $y$ is the leg index. Prior to the insertion of the hole, the density profiles are simply flat as the charge fluctuation is gapped in the Mott regime. With the insertion of a single hole by removing one electron (e.g., down spin electron) out, the hole distribution is extended over the whole system. Examples for the Hubbard model at $U/t=20$ and the $t$-$J$ model at $t/J=5$ are given in Fig. \ref{Fig:tJHubSim}. We keep up to $m=2048$ block states in the DMRG simulation with a negligible truncation of less than $10^{-10}$ and perform $100-500$ sweeps for decent convergence. Similar with the $t$-$J$ model, in the strong rung case such as $\alpha=0.5$, the hole density distribution is extended over the whole system, which is smooth and without charge modulation. In sharp contrast, the hole distribution $\left<n_x\right>$ develops a clear charge modulation in both systems in the isotropic limit $\alpha=1$. This clearly demonstrates the similarity of the two models and hence the correlated hopping terms are not crucial in determining ground state properties of a single doped hole in either model.

\begin{figure}[!htbp]
  \includegraphics[width=\linewidth]{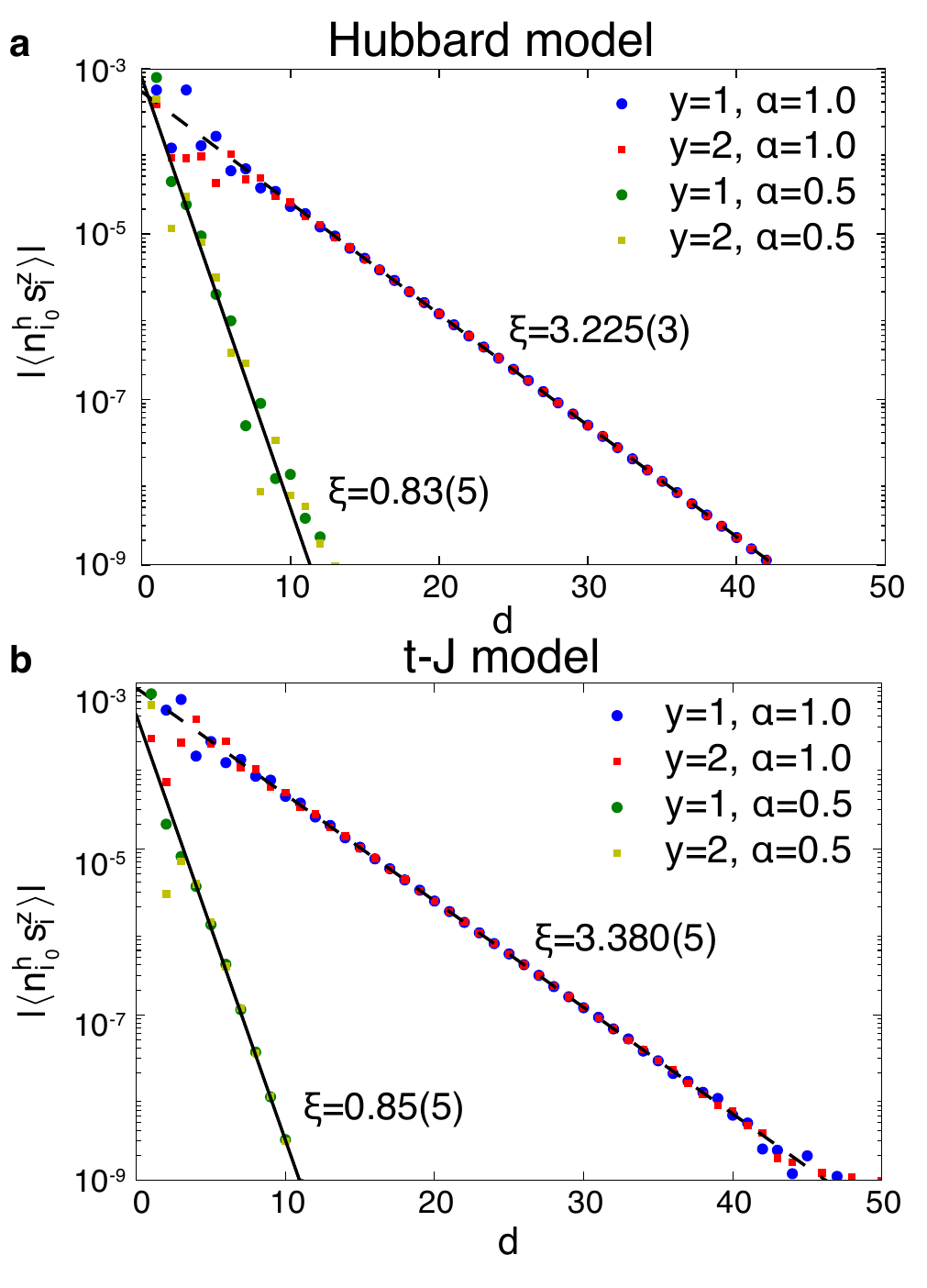}
  \caption{(Color online) Hole-spin correlation functions $|\left<n^h_{i_0}s^z_i\right>|$ for (a) $t$-$J$ model at $t/J=5$ and (b) Hubbard model at $U/t=20$. Here, $i_0=(50,2)$ is the hole site index, $i=(x,y)$ is the spin site index and $d=|x-50|$ is the distance between the hole and spin along the ladder. The exponentially decaying correlation functions show that the spin degrees of freedom are exponentially localized close to the dynamic hole for both $\alpha=0.5$ and $\alpha=1.0$. The solid lines show the linear fit to the data and $\xi$ is the spin-charge correlation length.}\label{Fig:SpinCharge}
\end{figure}

\paragraph{Spin-charge correlation:}%
Previous studies show that there is no spin-charge separation in the two-leg $t$-$J$ ladder for $0 \leq \alpha \leq 1$\cite{Zhu2015Quasiparticle,White2015TJ}. In this section, we show that this is also true for the two-leg Hubbard ladder. To prove this, we calculate the spin-charge correlation function $\left< n^h(i{_0}) s^z(i)\right>$ (see Fig. \ref{Fig:SpinCharge}) which measures the spin profile when a dynamic hole is on site $i_0=(50,2)$ and a spin is on site $i=(x,y)$ of a $N=100\times 2$ ladder. With the correlation function shown on a log scale as a function of distance $d=|x-50|$ along the ladder, the exponential confinement of the spin and charge is apparent in the linear $d$ dependence. A linear fit gives a decay length of $\xi=0.85(5)$ for the Hubbard model at $\alpha=0.5$ and $U/t=20$, showing that the spin and charge degrees of freedom are tightly bound together. A similar fit for $\alpha=1.0$ gives a length scale $\xi=3.380(5)$. For a direct comparison, we have also calculated the spin-charge correlation function for the $t$-$J$ model at $t/J=5$, which is related to the Hubbard coupling at $U/t=4t/J=20$. Consistent with the Hubbard model and previous studies, the spin-charge correlation function is also short-ranged with a correlation length $\xi=0.83(5)$ at $\alpha=0.5$ and $\xi=3.225(3)$ at $\alpha=1.0$. As the spin-charge correlation function is always short-ranged, we hence conclude that there is no spin-charge separation in both systems.

\begin{figure}
  \includegraphics[width=0.5\textwidth]{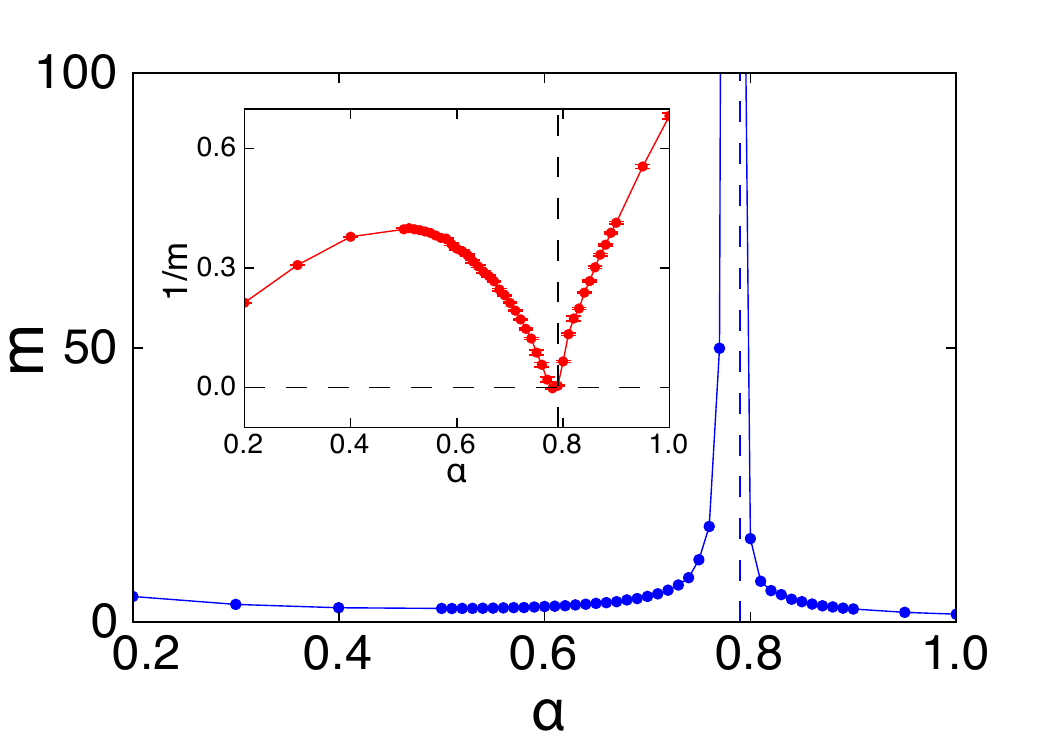}
  \caption{(Color online) The effective mass $m$ (in unit of $t$) of the single hole (or spin-charge object) of the two-leg Hubbard ladder at $U/t=20$. The effective mass diverges at $\alpha_c=0.79$, but remains finite in both $\alpha<\alpha_c$ and $\alpha>\alpha_c$ parameter regions. Inset shows the inverse of effective mass $1/m$ as a function of $\alpha$.}\label{Fig:EffectiveMass}
\end{figure}

\paragraph{Effective mass:}%
As the spin and charge of the hole are not separated, it is meaningful to ask whether the doped hole or the spin-charge bound particle behaves as a quasiparticle or is localized. If the doped hole behaves as a quasiparticle, we shall expect a finite effective mass $m$ which can be determined by the formula
\begin{equation}
  \Delta E_0(L)=E_{\mathrm{0}}^{\mathrm{one-hole}}(L) - E_{\mathrm{0}}^{\mathrm{half-fill}}(L) - \mathrm{const.}
  \label{Eq:EffectiveMass}
\end{equation}
Here $E_0^{\mathrm{half-fill}}(L)$ ($E_0^{\mathrm{one-hole}}(L)$) is the ground state energy of the system at hall-filling (with single doped hole), and $L$ is the length of the ladder. For a quasiparticle, $\Delta E(L)$ is expected to be proportional to $\pi^2/2mL^2$, where $m$ is the effective mass. On the contrary, if the injected hole is localized, $\Delta E(L)$ should decay exponentially with $L$ with a diverging effective mass. We find that $\Delta E(L)$ decays as $1/L^2$ in our simulation in both phases, which indicates that the doped hole seems not localized in real space. Plots of effective mass $m$ is shown in Fig. \ref{Fig:EffectiveMass}. As seen in the figure, $m$ is finite in both regions $\alpha<\alpha_c$ and $\alpha>\alpha_c$, while it diverges at the phase boundary between the two phases, e.g., $\alpha_c=0.79$ at $U/t=20$. These results are similar with previous studies of the $t$-$J$ model\cite{Zhu2013TJ,Zhu2015Charge, White2015TJ}, which further suggests that the simple $t$-$J$ model captures the ground state physics of a single doped hole in the Hubbard model.

\section{Elementary excitation energy in the two-leg $t$-$J$ ladder}\label{Sec:Elementary}
In the above section, we have shown that the single hole doped in the two-leg Hubbard ladder behaves qualitatively the same as in the two-leg $t$-$J$ ladder. Therefore, we will focus on the $t$-$J$ ladder in this section since it is much easier to simulate, whose results however should also apply for the Hubbard ladder. In addition to the ground state properties, it is also crucial to have a good understanding of the elementary excitations. In particular, calculating the energy gap to the (first and/or second) excited states is of fundamental importance for determining the intrinsic behavior of the single hole doped in the two-leg antiferromagnet, which is complementary to the study of ground state properties. For a direct comparison with previous studies\cite{Zhu2013TJ,Zhu2015Charge, White2015TJ}, we will focus on $t/J=3$ in what follows.

There is a standard way to find excited states and gaps using DMRG\cite{Stoudenmire2002}. First, we use DMRG to compute a ground state $|\psi_0\rangle$ of the Hamiltonian $H$ with energy $E_0$ in Eq. (\ref{Eq:TJMODEL}) to high accuracy. Then we define a new Hamiltonian $H_1=H+wP_0$, where $P_0=|\psi_0\rangle\langle \psi_0|$ is a projection operator and $w$ is an energy penalty for states not orthogonal to $|\psi_0\rangle$. If $w$ is large enough, the ground state $|\psi_1\rangle$ of $H_1$ with energy $E_1$ will be the second lowest eigenstate of $H$, i.e., the first excited state or a second ground state. Having found $|\psi_1\rangle$, we can continue to compute the next excited state $|\psi_2\rangle$ with energy $E_2$ if necessary by including both $P_0$ and $P_1=|\psi_1\rangle\langle \psi_1|$ in a new Hamiltonian $H_2=H_1+wP_1=H+wP_0+wP_1$.
For the current simulation, we use $w=100$ to make sure that different eigenstates are orthogonal to each other with a negligible overlap $|\langle \psi_i|\psi_j\rangle|^2\leq 10^{-13}$.

Practically, utilizing the above procedure requires a well-converged ground state. However, this is very hard for the single hole problem in general.\cite{Zhu2013TJ,Zhu2015Charge, White2015TJ} Although it is relatively easy to obtain a state with an extended hole density profile by performing a big number of (e.g.,  hundreds of) DMRG sweeps, this is still not enough to obtain the exact ground state of the system which is reflection symmetric, since the Hamiltonian itself has the reflection symmetry. To solve this problem, we will adopt the following symmetrization strategy: we first obtain a relatively converged ground state with an extended hole density profile, then we symmetrize the system by copying all the operators in the left part of the system to the right part, and use this as the initial state for the next step simulation. Generally, such a symmetrization process may raise the energy a  bit at the beginning, however, repeating this process several times will make the``initial'' state close enough to the exact ground state. Eventually, the real ground state of the system is obtained with reflection symmetric hole density distribution
and slightly lower ground state energy (e.g., $\sim 10^{-5}J$ for $N=140\times 2$).

\begin{figure}[!htbp]
\centering
\includegraphics[width=\linewidth]{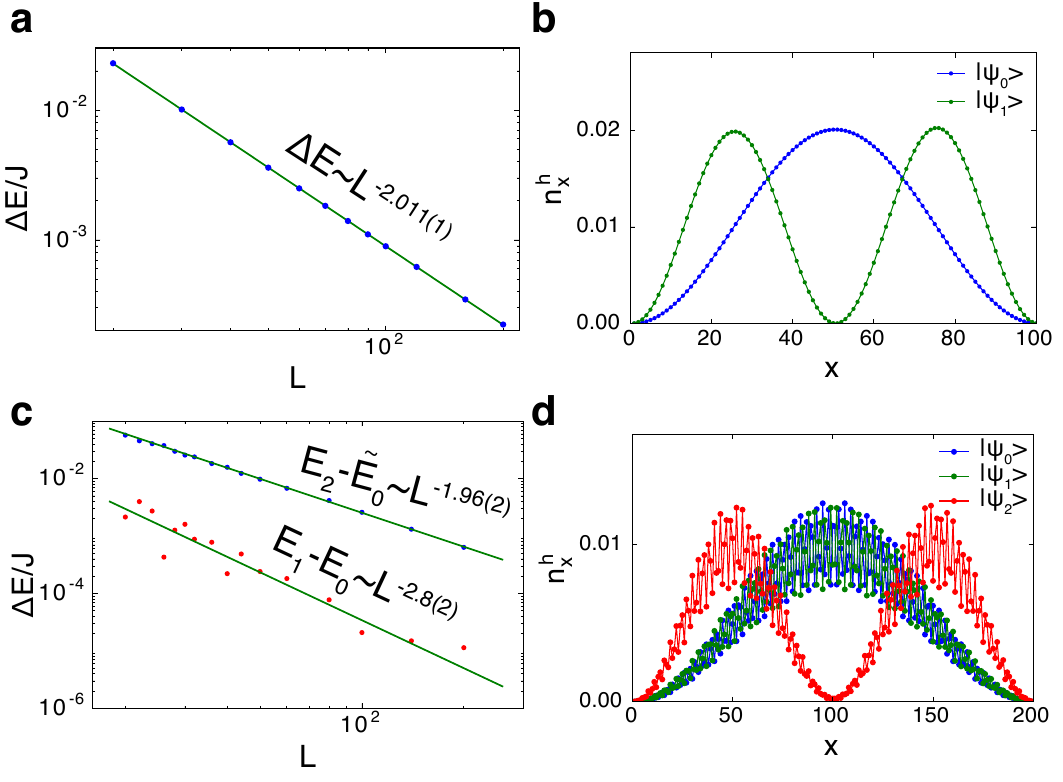}
\caption{(Color online) Excited state energy gaps and density profiles. (a) Scaling of  excitation energy $E_1 - E_0$ with lattice size at $\alpha = 0.5$. (b) Density profiles for ground state and first excited state at $\alpha = 0.5$. (c) Excitation energies $E_1 - E_0$ and $E_2 - \tilde{E}_0$ at $\alpha = 1.0$, where $\tilde{E}_0 = (E_1 + E_0) / 2$. (d) Density profiles for ground state, first excited state and second excited state at $\alpha=1.0$.}\label{Fig:Energy}
\end{figure}

With the exact ground state thus obtained, we continue to calculate the energies of excited states as described above. The results are given in Fig. \ref{Fig:Energy}. For $\alpha=0.5$, we find that the first excited state of the system $|\psi_1\rangle$ is consistent with the conventional quasiparticle picture. For example, the ground state $|\psi_0\rangle$ has a single peak while $|\psi_1\rangle$ has double peaks. Moreover, the elementary excitation energy $\delta E_1=E_1-E_0$ decays as $1/L^2$. These results are consistent with previous studies\cite{Zhu2013TJ,Zhu2015Charge, White2015TJ} and further establish the quasiparticle nature of the single doped hole in the strong rung limit $\alpha<\alpha_c$.

In contrast to $\alpha=0.5$ where the energy dispersion $\epsilon(k)$ is minimized at $k=0$, $\epsilon(k)$ is minimized at an incommensurate momentum $k=\pm k_0$ for $\alpha=1$, which gives rise to the oscillations in the charge density distribution (see Fig. \ref{Fig:Energy}(d)), as has been noted before \cite{Zhu2013TJ,Zhu2015Charge, White2015TJ}. Consequently, a``quasi-two-fold-degenerate'' ground state ($|\psi_0\rangle$ and $|\psi_1\rangle$) may be expected while $|\psi_2\rangle$ is the ``real'' first excited state, hence $E_2-E_1\gg E_1-E_0$. Indeed, our results are consistent with this and are plotted in Fig. \ref{Fig:Energy}(c). The energy splitting $E_1-E_0$ between the two ``quasi-degenerate'' ground states scales as $1/L^3$ (see Fig. \ref{Fig:Energy}(c)), which is caused by the combination of the charge modulation and open boundary condition. An explicit example can be found in the Supplementary Materials \ref{Sec:FreeFermion} for comparison. \footnote{Under periodic boundary condition, ground state $|\psi_0\protect\rangle$ and $|\psi_1\protect\rangle$ are exactly degenerate. Without the charge modulation, the degeneracy disappear and $|\psi_1\protect\rangle$ becomes the first excited state.}

In order to minimize the possible effect of ``quasi-degeneracy'', we define a ``proper'' ground state energy $\tilde{E}_0=(E_1+E_0)/2$ and the excitation energy gap $\Delta=E_2-\tilde{E}_0$. Similar with $\alpha=0.5$, we find that the energy gap $\Delta$ at $\alpha=1.0$ also decays as $1/L^2$. The hole density profile of the first excited state $|\psi_2\rangle$ shows double wavepackets in contrast to the single wavepacket of the ``quasi-degenerate'' ground states $|\psi_0\rangle$ and $|\psi_1\rangle$. This is also similar with $\alpha=0.5$ case. Our results hence suggests that the single hole at $\alpha=1.0$ also behaves as a quasiparticle when $L\gg\xi$ ($\xi$ denotes the spin-charge correlation length), which is consistent with previous studies. \cite{White2015TJ}

It is worth mentioning that although the doped hole behaves like a ``quasiparticle'' in both phases, there is a significant difference between them. In the strong-rung case ($\alpha<\alpha_c$), the spin and charge of the hole are tightly bound together without internal structure, so the hole behaves as a conventional Bloch quasiparticle. On the contrary, in the charge modulation phase $\alpha>\alpha_c$, the spin and charge of the hole are only loosely bound together with an interesting internal structure and nontrivial mutual statistics, which will lead to an important residual effect to be discussed in next section. This residual effect can dramatically change the local structure of the ground state wavefunction of the single hole, which however cannot be explained by a conventional quasiparticle picture.

\section{Residual Effect} \label{Sec:ResidualEffect}

\begin{figure}[!htbp]
\centering
\includegraphics[width=\linewidth]{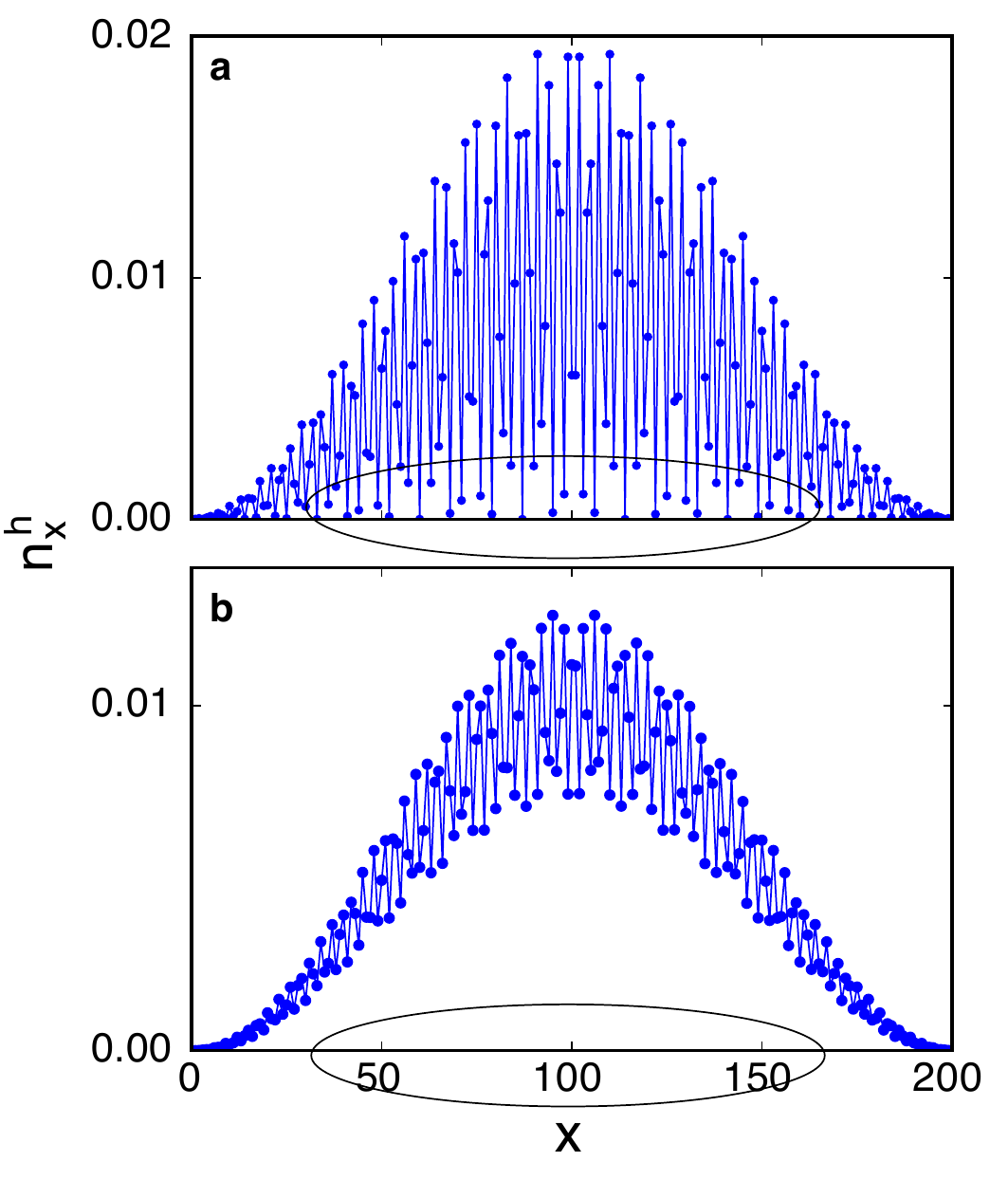}
\caption{(Color online) Ground state hole density profile $n_x^h$ for the (a) free fermion model in Section \ref{Fig:Single} and (b) isotropic two-leg $t$-$J$ ladder. It is clear that nodes are present at momentum $k\neq k_0$ for the free fermion system, while they are absent for the isotropic $t$-$J$ ladder.}\label{Fig:WaveFunction}.
\end{figure}

As just mentioned, although our results suggest that the single hole doped in the isotropic two-leg Hubbard and $t$-$J$ ladders  behaves as a quasiparticle, a mysterious residual effect is present which may not be explained by the conventional Bloch-quasiparticle picture. For a simple ``Bloch''-quasiparticle with energy dispersion located at incommensurate momentum $k=\pm k_0$, the ground state wavefunction is fast oscillating and crosses zeros (i.e., nodes) frequently at momentum $k\neq \pm k_0$, as labeled by the shaded region in Fig. \ref{Fig:WaveFunction}(a) for the free Fermion system. However, this is not true for the strongly interacting Hubbard and $t$-$J$ ladders. Although the hole density profile also shows significant modulation, the ground state wavefunction does not cross zeros at momentum $k\neq \pm k_0$, as seen in Fig. \ref{Fig:WaveFunction}(b). We argue that these nodes cannot be lifted by a simple finite-ranged Wannier function, suggesting that these nodes are unavoidable in a conventional quasiparticle picture.

A possible explanation is that the ground state wavefunction of the system consists two parts $|\psi_0\rangle = |\psi_0^L \rangle + |\psi_0^\xi \rangle$. Here $|\psi_0^L \rangle$ represents the long-wavelength contribution which accounts for the quasiparticle behavior of the single doped hole when the system size $L$ is much bigger than the spin-charge separation length scale $\xi$, i.e., $L\gg \xi$. Since there is no spin-charge separation, the system will only consider the doped hole as a single object while its internal structure is hidden. However, in the short length scale $\sim \xi$, the spin and charge of the doped hole will not behaves a whole object anymore since they are not tightly bound together. Instead, there is a nontrivial mutual statistics between them\cite{Weng1996}, i.e., a hole moving on the local antiferromagnetic spin background will induce a nontrivial phase-string effect. This has been shown to be relevant for the disappearance of the nodes structure in the two-leg $t$-$J$ ladder which was denoted as $|\psi_0^\xi\rangle$ here coming from the nontrivial mutual statistics between spin and charge part of the doped hole\cite{Wang2015VMC}. On the contrary, in the conventional quasiparticle picture, the spin and charge part of the doped hole are tightly bound together so that there is no internal structure.

\section{Conclusion}\label{Sec:Conclusion}
In this paper, we have systematically investigated the nature of a single hole doped in the two-leg antiferromagnet using large-scale DMRG simulation. We found that the doped hole in the Hubbard ladder behaves similarly with that in the $t$-$J$ ladder in the ground state. The elementary excitations of the doped hole are consistent with a quasiparticle. Interestingly, although the doped hole behaves like a quasiparticle in the long length limit, it is different with a simple Bloch-quasiparticle in the short length scale comparable with the spin-charge correlation length. In this limit, the nontrivial internal structure inside the loosely bound spin-charge object, namely the mutual statistics between the spin and charge of the doped hole, leads to a nontrivial residual effect dramatically changing the local structure of the ground state wavefunction. This may be potentially caused by the fundamental change of statistical sign structures as proposed in previous studies\cite{Weng2011Mott,Zhang2014Sign}. In the future, it will be important to design experiments to identify this nontrivial effect in other systems, which could potentially explain the role of sign structures in Hubbard and $t$-$J$ model directly.

\section{Acknowledgment}%
We thank Steven Kivelson, Xiaoliang Qi, Zheng-Yu Weng and Zheng Zhu for insightful discussions, and especially Zheng-Yu Weng for pointing out the novel residual effect. SL, HCJ and TPD were supported by the Department of Energy, Office of Science, Basic Energy Sciences, Materials Sciences and Engineering Division, under Contract DE-AC02-76SF00515.

\appendix
\renewcommand{\thefigure}{S\arabic{figure}}
\setcounter{figure}{0}
\renewcommand{\theequation}{S\arabic{equation}}
\setcounter{equation}{0}

\section{Free Fermion example} \label{Sec:FreeFermion}%

\begin{figure}[!htbp]
\centering
\includegraphics[width=\linewidth]{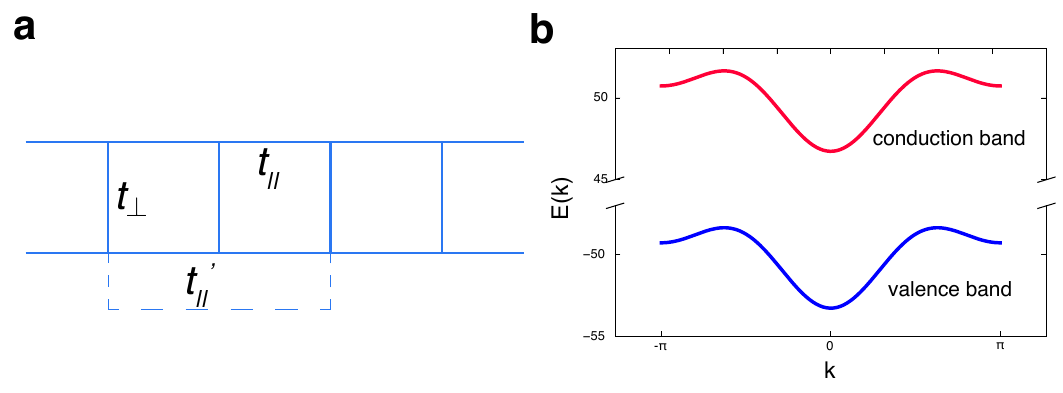}
\caption{(Color online) A noninteracting fermion model with energy minimum at non-zero momentum at half-filling under open boundary condition. (a) Coupling parameters and (b) Energy dispersion. With $t_{\parallel} = 1$, $t_{\parallel}^\prime = 0.632$ and $t_{\perp} = 50$, the energy minimum is at $k=\pm k_0=\pm 0.629\pi$.}\label{Fig:Single}
\end{figure}

In the previous study\cite{White2015TJ}, it has been argued that the single hole doped two-leg $t$-$J$ ladder can adiabatically connected to a non-interacting model, which has both first neighbor $t_{\parallel}$ and second neighbor $t_{\parallel}^\prime$ electron hopping along the ladder, and big enough hopping $t_{\perp}$ between ladders to open a band gap. When $\eta=t_{\parallel}^\prime/t_{\parallel} < 1/4$, the energy minimum is at $k=\pi$, while for $\eta>1/4$, the energy minimum is at $k=\pm k_0$. For a comparison with the two-leg single hole doped $t$-$J$ ladder, we directly calculate the non-interacting system with parameters as $t_{\parallel} = 1$, $t_{\parallel}^\prime = 0.632$ and $t_{\perp} = 50$, so that the non-interacting system is a band insulator at half-filling with energy minimum at finite momentum $k=\pm k_0$, whose energy dispersion is shown in Fig. \ref{Fig:Single}.

\begin{figure}[!htbp]
\centering
\includegraphics[width=\linewidth]{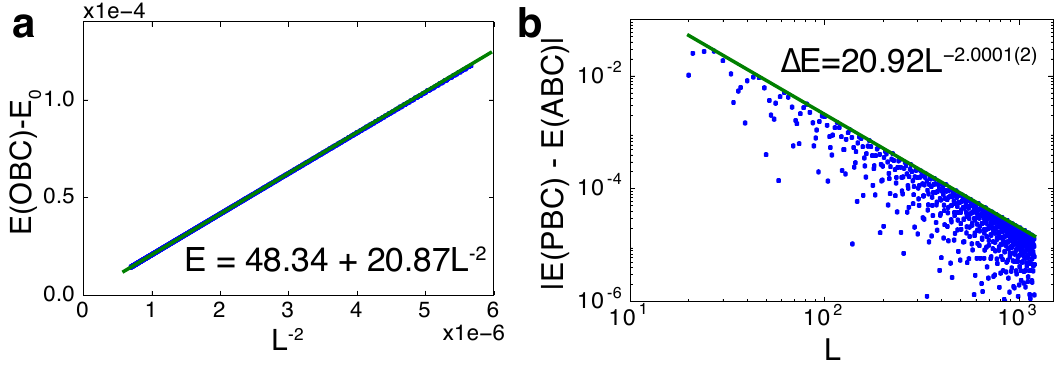}
\caption{(Color online) Effective mass calculated in two different ways. (a) $E_{\mathrm{OBC}} = \mathrm{const.} + \pi^2 / 2mL^2$ with $L=400 - 1200$. (b) $|E_{\mathrm{PBC}} - E_{\mathrm{ABC}}| = \pi^2 / 2mL^2$, whose decaying behavior is quite noisy and hard to make a clear conclusion for small lattice size. However, with large enough system size (e.g., $l=200 - 1200$), the overall envelop clearly decays as $1/L^2$. The effective mass determined by both ways are consistent with the exact value.} \label{Fig:EffmassSingle}
\end{figure}

The effective mass of the quasiparticle of the non-interacting model can be calculated theoretically as $m = 2t_{\parallel}^\prime/(16t_{\parallel}^{\prime 2} - t_{\parallel}^2) = 0.2345$. Similarly, we can also calculate $m$ in other ways, including (1) by computing the ground state energy of the system with open boundary condition (OBC) using Eq. (\ref{Eq:EffectiveMass}), and (2) by computing the energy difference of the system with periodic boundary condition (PBC) and anti-periodic boundary condition (ABC). As we can see in Fig. \ref{Fig:EffmassSingle}, both give us the same effective mass $m$ which are consistent the theoretical value.

\begin{figure}[!htbp]
\includegraphics[width=\linewidth]{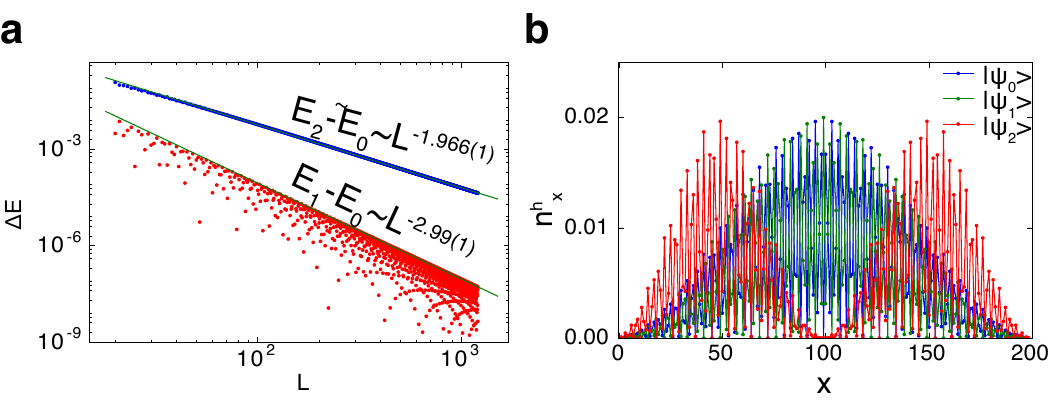}
  \caption{(Color online) (a) Excitation energy and (b) hole density profile for the non-interacting model in Ref.\cite{White2015TJ}. There are two quasi-degenerate ground states with energy splitting scaling as $1/L^3$ and similar hole density profile with single wavepacket, whereas the second excited state has a much higher energy where the energy difference between the excited state and the ground state is much larger which decays as $1/L^2$. Moreover, the hole density distribution has double wavepackets.}\label{Fig:ObcGap}
\end{figure}

Besides the ground state, we can also look into the excited states of the system, including both the first $E_1$ and second excited state $E_2$. The hole density distributions in $|\psi_0\rangle$ and $|\psi_1\rangle$ are similar with each other and both has a single wavepacket except a phase shift, indicating that they are (quasi) degenerate ground states. Instead, $|\psi_2\rangle$ is the first excited state with two wavepackets.  Interestingly, we find that the excitation energy of the single hole doped in the non-interacting system is qualitatively the same with the isotropic $t$-$J$ ladder. For example, the excitation energy scales as $E_2-E_0 \sim 1/L^2$, while the energy splitting scales as $E_1-E_0\sim 1/L^3$ in Fig. \ref{Fig:ObcGap}. Since the energy dispersion for both models has minimum at finite momentum $k=\pm k_0$ and both has open boundary condition, we therefore conclude the doped hole in the $t$-$J$ ladder also behaves as a quasiparticle.

\bibliographystyle{aipauth4-1}


\end{document}